\documentclass[aps,prb,reprint,amssymb,amsmath]{revtex4-2}
\usepackage{graphicx}
\usepackage{amsthm}
\usepackage{verbatim}

\def\be{\begin{equation}}
\def\ee{\end{equation}}
\def\ber{\begin{eqnarray}}
\def\eer{\end{eqnarray}}
\def\bern{\begin{eqnarray*}}
\def\eern{\end{eqnarray*}}

\def\rv{\mathbf{r}}

\def\0v{\mathbf{0}}
\def\1v{\mathbf{1}}
\def\2v{\mathbf{2}}
\def\3v{\mathbf{3}}

\def\pa{\partial}

\DeclareMathAlphabet\mathbfcal{OMS}{cmsy}{b}{n}

\def\Im{ {\rm Im} \, }

\begin{document}

\title{Exchange kernel $f^h_x(q,\omega)$ of electron liquid from the variational principle of  McLachlan}

\author{Vladimir~U.~Nazarov}
\affiliation{Fritz Haber Research Center for Molecular Dynamics and Institute of Chemistry, the  Hebrew University of Jerusalem, Jerusalem, Israel}
\email{vladimir.nazarov@mail.huji.ac.il}

\author{Vyacheslav M. Silkin}
\affiliation{Donostia International Physics Center (DIPC), Paseo de Manuel Lardizabal 4, E-20018 San Sebasti\'an, Spain}
\affiliation{Departamento de Pol\'imeros y Materiales Avanzados: Física, Qu\'imica y Tecnolog\'ia, Facultad de Ciencias Qu\'imicas, Universidad del País Vasco (UPV-EHU), Apdo. 1072, E-20080 San Sebastián, Spain}
\affiliation{IKERBASQUE, Basque Foundation for Science, 48011 Bilbao, Spain}

\begin{abstract}
By minimizing, in the $L_2$ norm, the difference between the left- and the right-hand sides of the time-dependent Schr\"{o}dinger equation,
the variational principle of McLachlan (McLVP) [A.~McLachlan, Molecular Physics {\bf 8}, 39 (1964)] provides a powerful tool for the generation of equations of motion.
If the trial wave function  is the Slater determinant, McLVP produces a temporally and spatially nonlocal  exchange potential [V.~U.~Nazarov, Phys. Rev. B {\bf 87}, 165125 (2013)]. We study the performance of the corresponding wave-vector and frequency-dependent exchange kernel $f^h_x(q,\omega)$  of the homogeneous electron liquid. While the McLVP-based $f^h_x(q,\omega)$ lacks correlations by construction, we find that it accurately accounts for exchange, reproducing features in the quantum Monte Carlo data, which the known constraint-based kernels miss.
We argue that the complementary use of the McLVP- and the constraint-based exchange-correlation kernels  will enhance the performance of the linear response time-dependent density functional theory of the electron liquid.
\end{abstract}

\maketitle

\section{Introduction}

A concept of the exchange-correlation (xc) kernel $f_{xc}(\rv,\rv',\omega)$ is central to the linear-response time-dependent density 
functional theory (TDDFT) \cite{Gross-85}. This quantity serves to relate the interacting-electron density response function $\chi(\rv,\rv',\omega)$ to its  
independent-particle  [Kohn-Sham (KS)] counterpart $\chi_s(\rv,\rv',\omega)$, doing this via the equality 
(we use atomic units throughout)
\begin{equation}
\chi^{-1}(\rv,\rv',\omega)=\chi^{-1}_s(\rv,\rv',\omega)-\frac{4\pi}{|\rv-\rv'|} -f_{xc}(\rv,\rv',\omega).
\label{ks}
\end{equation}
For an arbitrary  quantum-mechanical system, $\chi_s$ can be explicitly and exactly  written in terms of the single-particle orbitals and their corresponding eigenvalues, while the determination of $\chi$ via Eq.~\eqref{ks} requires the construction of approximations to $f_{xc}$
\cite{Giuliani&Vignale}.

A specific, but fundamentally important  case is the xc kernel of the homogeneous electron gas (HEG). Due to the translational symmetry, in this case, the kernel can be conveniently Fourier transformed to the wave-vector variable $q$. Over years, much effort has been invested in the construction of $f_{xc}(q,\omega)$, starting from the long-wave limit $f_{xc}(q\to 0,\omega)$ \cite{Gross-85,Iwamoto-87,Qian-02}, and later including the spatial non-locality via finite $q$ \cite{Holas-79,Constantin-07,Nazarov-13-2,Bates-16,Panholzer-18,Ruzsinszky-20,Nepal-21,Kaplan-22,Kaplan-23}.

Existing approximations to $f_{xc}(q,\omega)$ can be classified into the groups of the constraint-based (CB) and {\it ab initio} ones. The former use the known exact limiting properties of the kernel, while interpolating for the intermediate values of the $q$ and $\omega$ arguments. Examples are kernels of Refs.~\onlinecite{Constantin-07,Bates-16,Panholzer-18,Ruzsinszky-20,Nepal-21,Kaplan-22,Kaplan-23}. The latter category derive the kernel, essentially adverting to the many-body  time-dependent Schr\"{o}dinger equation \cite{Holas-79,Richardson-94,Nazarov-13-2}.

Unless  a CB kernel is based on the best fit to specific simulation data, the interpolation used in its construction is largely arbitrary. As a result, as we show in this paper, important features of the wave-vector and frequency dependence of the kernel may be lost. We test the CB kernels against the quantum Monte Carlo (QMC) (essentially exact) simulations results,  finding significant discrepancies  in the range of the wave-vector $k_F \lesssim 2 k_F$, where $k_F$ is the Fermi radius.
By comparing to the {\it ab initio} exchange only kernel based on the time-dependent variational principle of  McLachlan (McLVP) \cite{McLachlan-64}, we prove that the said discrepancies originate from the inaccuracy of the CB kernels in their treating exchange.
Finally, we propose that the complementary use of the McLVP-based kernel and the CB ones can lead to the overall improvement of the theory.

The structure of this paper is the following. In Sec.~\ref{ML}, we briefly remind of the variational principle of  McLachlan and of its relation to 
the problem of self-consistent field equations. In Sec~\ref{RD}, we present results of calculations of the McLVP-based exchange kernel, give their interpretation, and conduct comparison with  the CB kernels and with the QMC simulations data.
Conclusions are collected in Sec.~\ref{CN}.

\section{Background: M\lowercase{c}L\lowercase{achlan's} variational principle, exchange potential, and exchange kernel of HEG}
\label{ML}

The time-dependent (TD) variational principle of McLachlan  \cite{McLachlan-64} minimizes the functional
\begin{equation}
F(t)=\int \left| \left[ i \frac{\pa}{\pa t} - \hat{H}(t)\right] \Psi(\rv_1,\dots,\rv_N,t) \right|^2 d\rv_1 \dots d\rv_N,
\label{VP}
\end{equation}
where $\Psi(\rv_1,\dots,\rv_N,t)$ is a trial wave-function of an $N$-body system, and $\hat{H}(t)$ is, generally speaking TD, Hamiltonian.
In Eq.~\eqref{VP}, only the time-derivative of $\Psi$ is varied, while $\Psi$ itself is considered known at the time $t$, which procedure determines the time-propagation.

If the time-derivative of $\Psi$ is varied unrestrictedly, then McLVP, obviously,  recovers the original TD Schr\"{o}dinger equation.
If $\Psi$ is the Slater determinant and time-derivatives of the single-particle {\it orbitals} are the varied quantities, then this variational principle  yields  TD Hartree-Fock equations \cite{Nazarov-85}.

\subsection{Exchange potential}

Alternatively, in the context of TDDFT, if $\Psi$ is the Slater determinant, but all the orbitals obey the single-particle 
Schr\"{o}dinger equation
\begin{equation}
i \frac{\pa \phi_\alpha(\rv,t)}{\pa t}= \left[-\frac{1}{2} \Delta + v_s(\rv,t)\right] \phi_\alpha(\rv,t)
\end{equation}
with the same multiplicative potential $v_s(\rv,t)$, and the time-derivative of the latter potential is the varied quantity, then one arrives at the equation for the exchange potential \cite{Nazarov-13-2}
\begin{equation}
\begin{split}
n(\rv,t) v_x(\rv,t)=
 \int  \left[ v_x(\rv',t)
\! - \!  \frac{1}{|\rv-\rv'|} \right] 
|\rho(\rv,\rv',t)|^2
d\rv'  \\
 +  \int    
 \frac{\rho(\rv,\rv',t)\rho(\rv',\rv'',t)\rho(\rv'',\rv,t)}{|\rv'-\rv''|}  
d\rv'  d\rv'',
\end{split}
\label{main4}
\end{equation}
where
$v_x(\rv,t)=v_s(\rv,t)-v_{ext}(\rv,t)-v_H(\rv,t)$,  $v_{ext}(\rv,t)$ and $v_H(\rv,t)$ are the external and the Hartree potentials, respectively, 
and
\begin{align}
&n(\rv,t)=\sum\limits_{\alpha=1}^N |\psi_\alpha(\rv,t)|^2,\\
&\rho(\rv,\rv',t)=\sum\limits_{\alpha=1}^{N} \psi_\alpha(\rv,t) \psi_\alpha^*(\rv',t) 
\label{dm}
\end{align}
are the particle density and the single-particle density-matrix, respectively
\footnote{(I) In the static setup, Eq.~\eqref{main4} had been previously obtained by an alternative method and is known as an equation for the Localized Hartree-Fock potential \cite{Sala-01}; (II) When working on Refs.~\onlinecite{Nazarov-85,Nazarov-13-2}, we were not aware of Ref.~\onlinecite{McLachlan-64}}.

\subsection{Exchange kernel}

With the use of Eq.~\eqref{main4} and of the definition
\begin{equation}
f_x(\rv,\rv',t-t')=\frac{\delta v_x(\rv,t)}{\delta n(\rv',t')},
\label{fd}
\end{equation}
the  McLVP-based exchange kernel of HEG $f^h_x(q,\omega)$ was constructed in Ref.~\onlinecite{Nazarov-13-2}. 

We note that (I) The McLVP-based potential  defined by Eqs.~\eqref{main4}-\eqref{dm} is local in time w.r.t. the {\it orbitals}, but it is nonlocal  w.r.t. the density. This property results in the exchange kernel obtained through Eq.~\eqref{fd} being, in general, temporally nonlocal, i.e.,  to exhibit the frequency dependence;
(II) The McLVP-based potential is an advanced exchange only potential, in particular, it is free of self-interaction and it supports the derivative discontinuity \cite{Perdew-82} in the energy dependence on the fractional particle number \cite{Nazarov-15-2};
(III) The McLVP-based potential presents an efficient alternative to the optimized effective potential (OEP) \cite{Sharp-53,Talman-76,Petersilka-96},
since the solution of Eq.~\eqref{main4} is incomparably easier than the solution of the OEP equation;
(IV) The McLVP-based $f^h_x(q,\omega)$ is non-singular, causal, and it  satisfies the requirement of the positivity of dissipation, 
all of which properties are violated by the first-order perturbation theory \cite{Holas-79,Giuliani&Vignale}.

Our numerical results for the McLVP $f^h_x(q,\omega)$  below are based on Eqs.~(12)-(16) and Appendix C of Ref.~\onlinecite{Nazarov-13-2}.

\section{Results and discussion}
\label{RD}

\subsection{Static regime}

\begin{figure*}[h!] 
\hspace{-1 cm}
\includegraphics[width= 0.75 \textwidth, trim=35 0 0 0, clip=true]{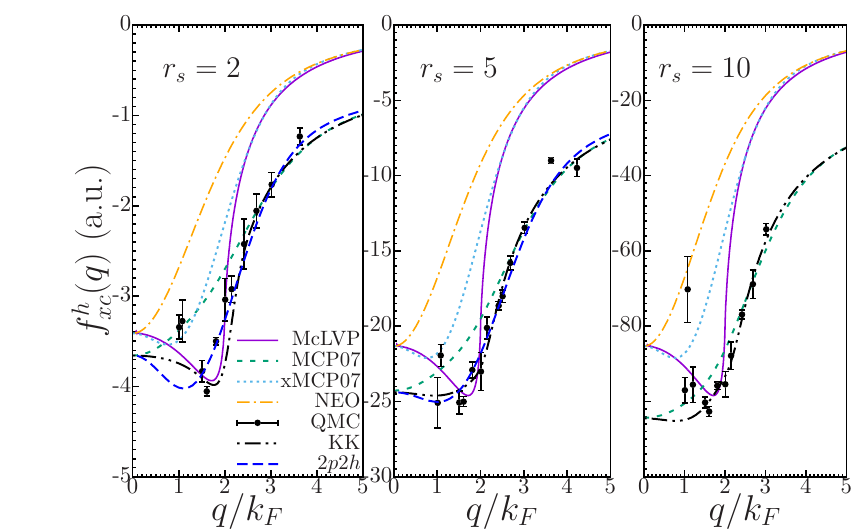} 
\caption{\label{stat}
Static exchange-correlation kernel of HEG of $r_s=2$, $5$, and $10$.
Solid (magenta) lines are McLVP exchange  kernel of Ref.~\onlinecite{Nazarov-13-2}.
Short dashed (green) lines are the modified Constantin-Pitarke (MCP07) xc kernel of Ref.~\onlinecite{Ruzsinszky-20}.
Dotted (light blue) lines are the exchange-only version of the latter (xMCP07).
Dash-dotted (orange) lines are nonlocal energy optimized (NEO) exchange only kernel of Ref.~\onlinecite{Bates-16}.
Dash-dotted-dotted (black) lines   are the QMC data-adjusted interpolation of Ref.~\onlinecite{Kaplan-23} (KK).
Symbols with error bars are $f^h_{xc}(q)$ by QMC simulations of Ref. \onlinecite{Moroni-95}.
Long dashed (blue) lines are the two-particle-two-hole $f_{xc}$ calculations of Ref.~\onlinecite{Panholzer-18} ($2p2h$)
(there are no $2p2h$ data available for $r_s=10$).
}
\end{figure*}

In Fig.~\ref{stat}, we plot the static kernels ($\omega=0$)  of the  HEG of  the three different densities, using the essentially exact QMC data of Ref.~\onlinecite{Moroni-95},
the recent  CB Modified Constantin-Pitarke (MCP07) xc kernel \cite{Ruzsinszky-20}, the exchange-only version of the latter (xMCP07), 
the Nonlocal Energy Optimized  (NEO) exchange kernel \cite{Bates-16}, 
the McLVP-based exchange kernel \cite{Nazarov-13-2}, the CB xc kernel of 
Kaplan and Kukkonen (KK) \cite{Kaplan-23}, and the two-particle-two-hole xc kernel of Panholzer {\it et al.} ($2p2h$) \cite{Panholzer-18}. The two latter xc kernels use QMC data in their construction.

At the considered electron densities, at larger values of the wave-vector ($q>2 k_f$), MCP07, KK, and $2p2h$, but not McLVP, xMCP07, or NEO, compare well with the QMC data, suggesting that correlations play a significant role in that $q$-domain. However, the QMC data have a prominent dip within the interval $k_F \lesssim q  \lesssim 2 k_F$, which dip the MCP07 kernel, both with and without
account of correlations, ignores completely. 
On the contrary, the McLVP kernel reproduces this feature fairly well. 
The KK compares well with QMC data everywhere, but it should be remembered that this kernel involves the best fit to the QMC data.
At $r_s=2$, the position of the dip in the  McLVP and KK results agree with that of the QMC data, suggesting that the two latter approximations perform well at this density and within this $q$-range. The dip is, however, shifted to the lower $q$-values for the $2p2h$ xc kernel.
At $r_s=5$, the $2p2h$ and KK agree between themselves, while McLVP exhibits a deeper dip, which is shifted to higher $q$-s. The sparsely scattered QMC data with wide error bars do not permit to conclusively judge which results are more accurate in this case.
At $r_s=10$, when the $2p2h$ results are not available, McLVP exhibits a pronounced dip, while KK demonstrates a shallow one positioned at lower $q$-s. The QMC data, seemingly, favour the McLVP prediction in this case.

The following conclusions can be drawn from the above observations: 
\begin{enumerate} 
\item
The dip in  question is due to exchange, not to correlations; 
\item 
In the wave-vector range $k_F \lesssim q  \lesssim 2 k_F$,  MCP07 does not reproduce exchange well, but McLVP and KK do;
\item 
In that $q$-range, correlations are of lesser importance.
\end{enumerate}

Regarding the xMCP07 and NEO kernels, they satisfy the limiting conditions necessary for exchange at $q\to 0$ and $q\to\infty$, but, otherwise, they perform poorly in between.

\begin{figure*}[h!] 
\hspace{-1 cm}
\includegraphics[width= 0.75 \textwidth, trim=35 0 0 0, clip=true]{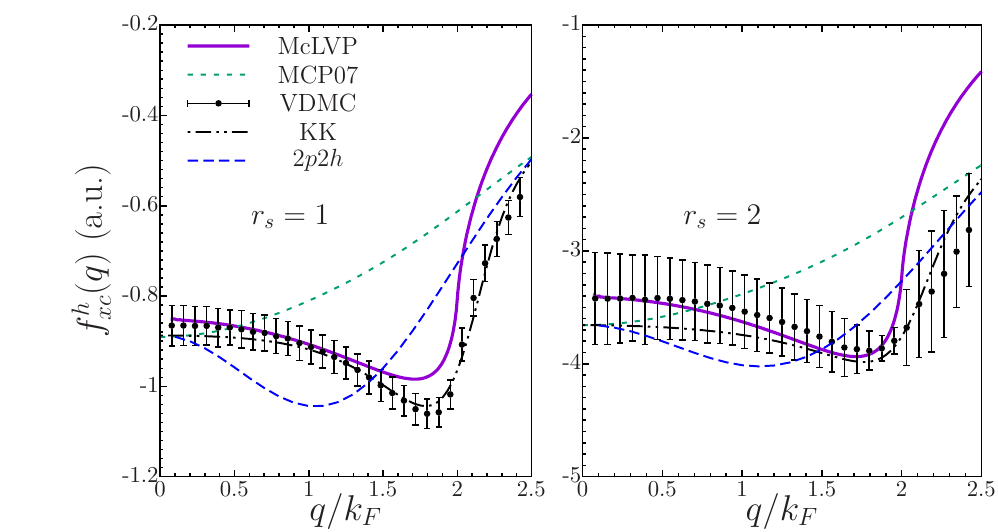} 
\caption{\label{stat12}
Static exchange-correlation kernel of HEG of $r_s=1$ and $2$.
Solid (magenta) lines are McLVP exchange  kernel of Ref.~\onlinecite{Nazarov-13-2}.
Short dashed (green) lines are the MCP07 xc kernel of Ref.~\onlinecite{Ruzsinszky-20}.
Symbols with error bars are $f^h_{xc}(q)$ by the variational diagrammatic Monte Carlo (VDMC) simulations of Ref. \onlinecite{Kukkonen-21}.
Dash-dotted-dotted (black) lines are the VDMC-fitted interpolation of Ref.~\onlinecite{Kaplan-23} (KK).
Long dashed (blue) lines are the two-particle-two-hole $f_{xc}$ calculations of Ref.~\onlinecite{Panholzer-18} ($2p2h$).
}
\end{figure*}

The older QMC data of Ref.~\onlinecite{Moroni-95} we use in Fig.~\ref{stat} do not include the region $q\le k_F$. In Fig.~\ref{stat12}, we test the performance of the McLVP, MCP07, and $2p2h$ kernels against the recent variational diagrammatic Monte Carlo (VDMC) simulations data \cite{Kukkonen-21} and their KK interpolation \cite{Kaplan-23} for $r_s=1$ and $2$.
The overall agreement between McLVP, VDMC, and KK is good in the whole range of $q \le 2 k_F$, while $2p2h$, and especially MCP07, demonstrate a qualitative disagreement with the former kernels at $k_F\le q \le 2 k_F$. 

\subsection{Dynamic regime}

\begin{figure*}[h!] 
\hspace{-1 cm}
\includegraphics[width= 0.75 \textwidth, trim=73 0 0 0, clip=true]{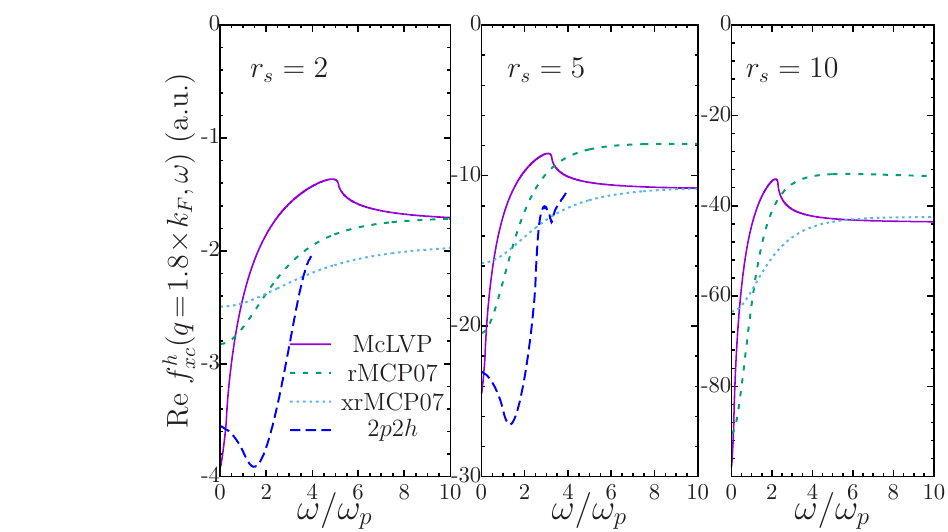} 
\caption{\label{dynr}
Real part of the dynamic exchange-correlation kernel of HEG of $r_s=2$, $5$, and $10$.
Solid (magenta) lines are McLVP exchange kernel of Ref.~\onlinecite{Nazarov-13-2}.
Dashed (green) lines are the revised MCP07 (rMCP07) xc kernel of Ref.~\onlinecite{Kaplan-22}.
Dotted (light blue) lines are the exchange-only version of the latter (xrMCP07).
Long dashed (blue) lines are the two-particle-two-hole $f_{xc}$ calculations of Ref.~\onlinecite{Panholzer-18} ($2p2h$)
The wave-vector is set at the $q=1.8 \times k_F$ value.
}
\end{figure*}

In Fig.~\ref{dynr}, we plot and compare the frequency-dependence of the real parts of the McLVP, the revised MCP07 (rMCP07) \cite{Kaplan-22}  kernels, the exchange-only version of the latter (xrMCP07), and the $2p2h$ xc kernel
\footnote{MCP07 and rMCP07 kernels differ in the dynamic regime only.}.
The wave-vector is set at $q=1.8\times k_F$, which is within the range where the McLVP kernel is superior in the static regime (see Fig.~\ref{stat}). 
As in the static case, the real part of the dynamic  rMCP07 kernel is structureless, which is not surprising, considering a smooth interpolation used in the construction of the latter. On the contrary, the frequency dependence of the real part of the dynamic McLVP kernel exhibits a hump at the edge of the particle-hole excitation continuum.

Although the shape of the $\omega$-dependence of the $2p2h$ kernel differs largely from that of McLVP, 
the singularity at the particle-hole continuum edge is clearly visible in the $2p2h$ data at $r_s=5$ as well.
At $r_s=2$, the singularity is located at $\omega>4 \omega_p$, where the $2p2h$ data do not extend to, while for $r_s=10$ there is no $2p2h$ data available.

\begin{figure*}[h!] 
\hspace{-1 cm}
\includegraphics[width= 0.75 \textwidth, trim=73 0 0 0, clip=true]{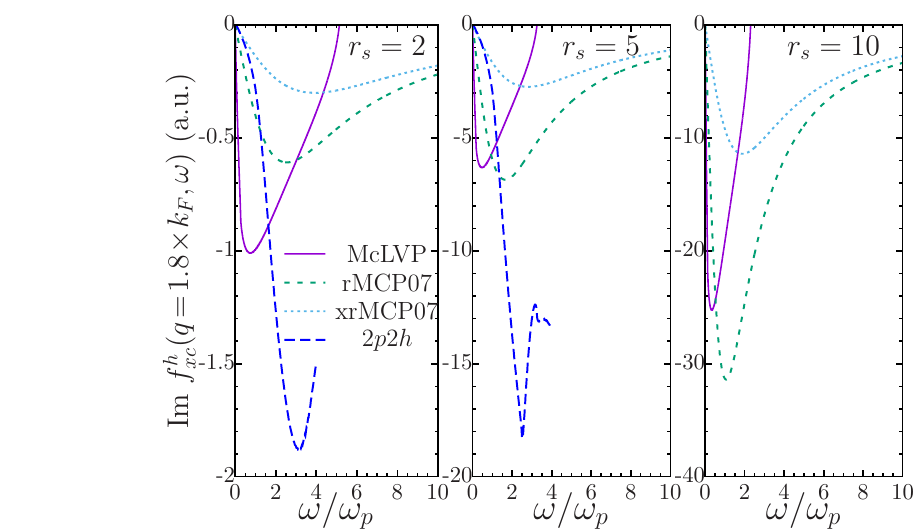} 
\caption{\label{dyni}
Same as Fig.~\ref{dynr}, but for the imaginary parts of the dynamic x(c) kernels.
}
\end{figure*}

In Fig.~\ref{dyni}, imaginary parts of the McLVP, rMCP07, and $2p2h$ kernels are plotted. We note that, for McLVP, $\Im f^h_x(q,\omega)$ is zero outside the particle-hole excitation continuum, which is a common property of the {\it ab initio} exchange-only kernels \cite{Holas-79,Giuliani&Vignale}. As discussed above with respect to the real parts of the kernels, the edge of the particle-hole continuum displays itself as a singularity in the $2p2h$ kernel.

The dielectric function $\epsilon(q,\omega)$  is related to the density response function $\chi(q,\omega)$ by the equality
\cite{Giuliani&Vignale}
\begin{equation}
\frac{1}{\epsilon(q,\omega)}=1+\frac{4\pi}{q^2} \chi(q,\omega),
\end{equation}
whereas \cite{Gross-85}
\begin{equation}
\frac{1}{\chi(q,\omega)}=\frac{1}{\chi_s(q,\omega)} -\frac{4\pi}{q^2} -f_{xc}(q,\omega).
\end{equation}

\begin{figure*}[h!] 
\hspace{-1 cm}
\includegraphics[width= 0.75 \textwidth, trim=73 0 0 0, clip=true]{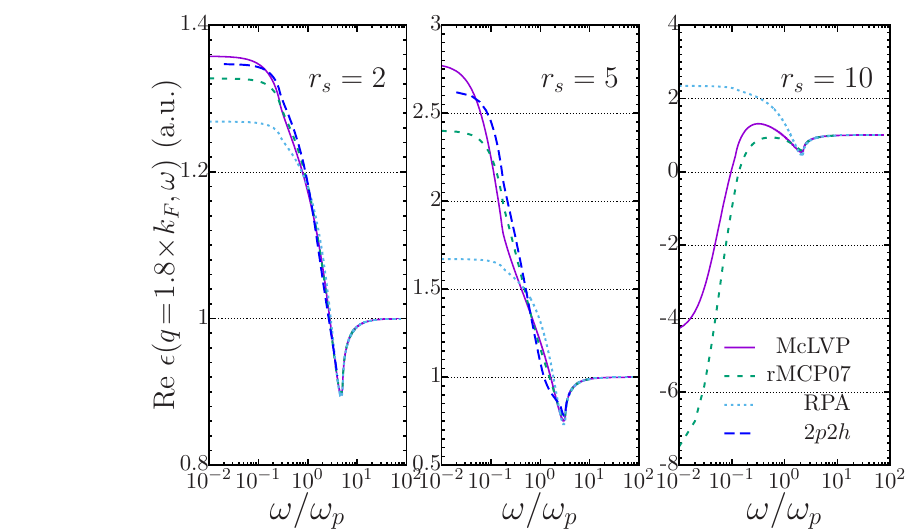} 
\caption{\label{epsr}
Real part of the dielectric function of HEG of $r_s=2$, $5$, and $10$.
Solid (magenta) lines are McLVP-based dielectric function of Ref.~\onlinecite{Nazarov-13-2}.
Dashed (green) lines are the rMCP07-based dielectric function.
Dotted (light blue) lines are the RPA [$f^h_{xc}(q,\omega)=0$] dielectric function.
Long dashed (blue) lines are the $2p2h$ results of Ref.~\onlinecite{Panholzer-18} 
The wave-vector is set at the $q=1.8 \times k_F$ value.
In order to well resolve the low-frequency behaviour of $\epsilon(q,\omega)$, the logarithmic scale is applied to the $\omega$-axis.
}
\end{figure*}

\begin{figure*}[h!] 
\hspace{-1 cm}
\includegraphics[width= 0.75 \textwidth, trim=73 0 0 0, clip=true]{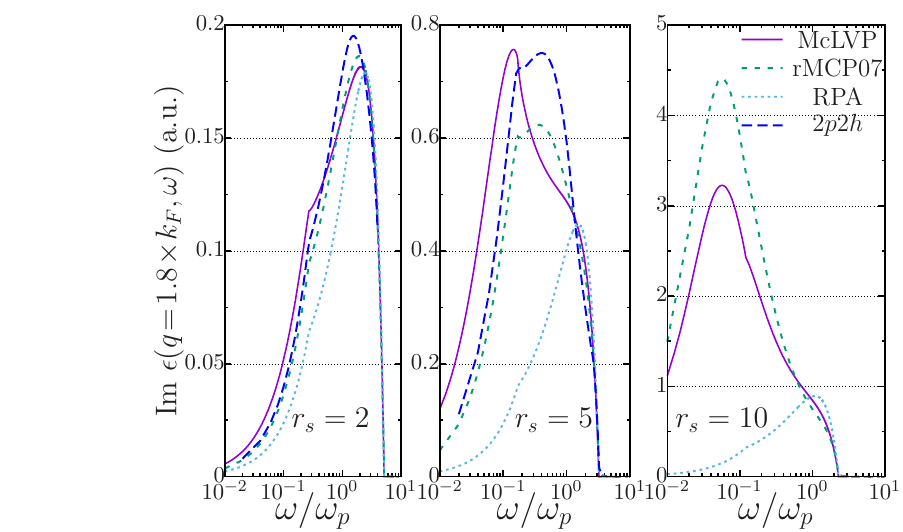} 
\caption{\label{epsi}
Same as Fig.~\ref{epsr}, but for the imaginary part of the dielectric function.
}
\end{figure*}

\begin{figure*}[h!] 
\hspace{-1 cm}
\includegraphics[width= 0.75 \textwidth, trim=73 0 0 0, clip=true]{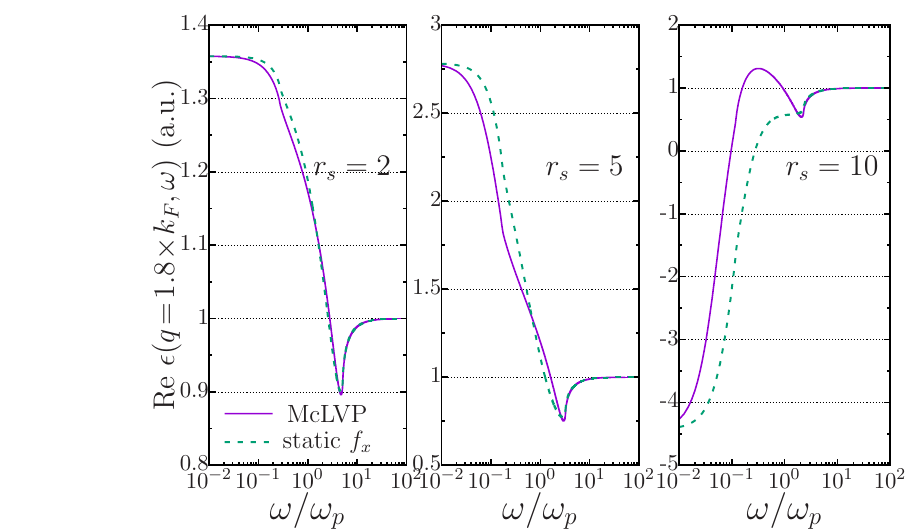} 
\caption{\label{epsstr}
Real part of the dielectric function of HEG of $r_s=2$, $5$, and $10$.
Solid (magenta) lines are the McLVP results.
Dashed (green) lines are the dielectric function obtained with the static version [$f^h_x(q,\omega=0)$] of the same kernel.
The wave-vector is set at the $q=1.8 \times k_F$ value.
}
\end{figure*}

\begin{figure*}[h!] 
\hspace{-1 cm}
\includegraphics[width= 0.75 \textwidth, trim=73 0 0 0, clip=true]{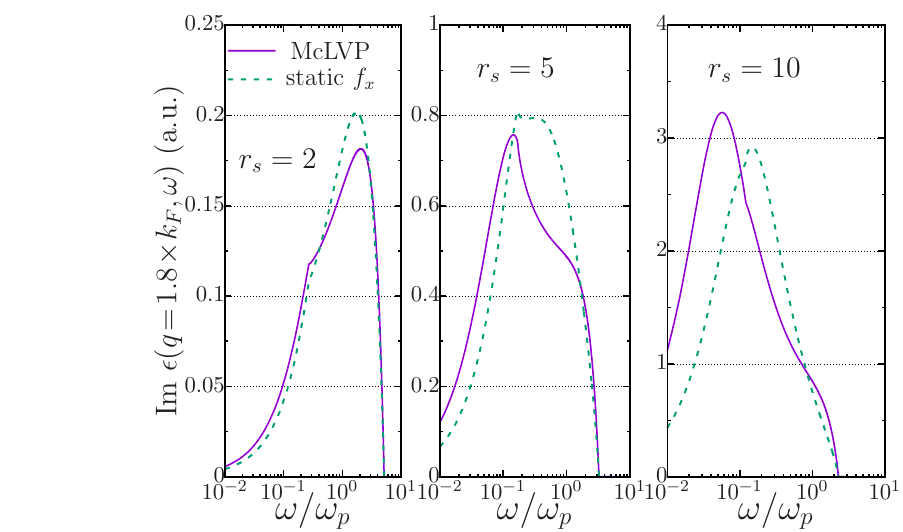} 
\caption{\label{epssti}
Same as Fig.~\ref{epsstr}, but for the imaginary part of the dielectric function.
}
\end{figure*}

In Figs.~\ref{epsr} and \ref{epsi}, we plot the real and imaginary parts, respectively, of $\epsilon(q,\omega)$, using the McLVP exchange kernel, rMCP07 xc kernel, $2p2h$ xc kernel, and the random phase approximation (RPA) [$f^h_{xc}(q,\omega)=0$]. We observe that xc affects mostly the lower-frequency part of the dielectric function.
Estimating the role of the frequency dependence, in Figs.~\ref{epsstr} and \ref{epssti} we compare $\epsilon(q,\omega)$ obtained with the use of the McLVP kernel versus the static version of the latter [$f^h_x(q,\omega=0)$]. Again, we find that the frequency dependence of the kernels is only important in the lower-$\omega$ part of the spectra.

\section{Conclusions}
\label{CN}

Based on the time-dependent variational principle of McLachlan, we have conducted calculations of the wave-vector and frequency-dependent exchange kernel of the homogeneous electron gas.
We have compared results with those of the constrain-based exchange-correlation kernels (r)MCP07, the two-particle-two-hole exchange-correlation kernel ($2p2h$), with the QMC simulation data, and their interpolation. 

While the McLVP kernel misses correlations altogether, we have found it very accurate in accounting for exchange. In particular, in the static regime ($\omega=0$), comparing with the QMC data, we have unambiguously identified a prominent structure in the wave-vector dependence of the kernel as being due to exchange. 

In the dynamic regime, McLVP yields structured spectra as well, while rMCP07 produces a smooth structureless dependence of $f^h_{xc}(q,\omega)$ on $\omega$.
We argue, that the smoothness of the rMCP07 kernel originates from the  interpolation between the exactly known limiting values, which interpolation is unable to catch the non-monotonic features located in between. On the contrary, being an {\it ab initio}, although  the exchange only kernel, the McLVP-based one successfully resolves structures, provided they are due to exchange. Comparison with the QMC-based $2p2h$ kernel gives an additional support to our interpretation.

We expect, that  the complementary use of the McLVP kernel, which accurately handles exchange, and the constraint-based kernels, which are better fit for treating correlations, will result in the advancement of the linear response TDDFT. Finally, since  the McLVP kernel is generated with  a variational principle, an avenue to the inclusion of correlations  by means of using the trial wave-functions other than a single Slater determinant is clearly foreseen. This work is underway. 

Fortran 90 code implementing the calculation of the McLVP-based exchange kernel is available from authors upon request.

\acknowledgments

We are indebted to Carl A. Kukkonen and Kun Chen for providing us with their original variational diagrammatic Monte Carlo calculations data.
V.U.N. gratefully acknowledges the hospitality of the Donostia International Physics Center.


%

\end{document}